\documentclass[twocolumn]{aastex63} 
\usepackage{amsmath,  amssymb, url, graphicx}
\usepackage{balance}

\renewcommand{\vec}[1]{\pmb{#1}}

\newcommand{\beq}{\begin{equation}}
\newcommand{\eeq}{\end{equation}}

\newcommand{\T}{{\cal T}}

\newcommand{\M}{{\cal M}}

\shorttitle{Giant Hall Waves in Pulsars}

\begin{document}
\title{Giant Hall Waves Launched by Superconducting Phase Transition in Pulsars}

\correspondingauthor{Ashley Bransgrove}
\email{abransgrove@princeton.edu}

\author{Ashley Bransgrove} 
\affil{Princeton Center for Theoretical Science and Department of Astrophysical Sciences, Princeton University, Princeton, NJ 08544, USA}
\affil{Physics Department and Columbia Astrophysics Laboratory, Columbia University, 538 West 120th Street, New York, NY 10027}

\author{Yuri Levin}
\affil{Physics Department and Columbia Astrophysics Laboratory, Columbia University, 538 West 120th Street, New York, NY 10027}
\affil{Center for Computational Astrophysics, Flatiron Institute, 162 5th Avenue, 6th floor, New York, NY 10010
}
\affil{Department of Physics and Astronomy, Monash University, Clayton, VIC 3800, Australia}

\author{Andrei M. Beloborodov}
\affil{Physics Department and Columbia Astrophysics Laboratory, Columbia University, 538 West 120th Street, New York, NY 10027}
\affiliation{Max Planck Institute for Astrophysics, Karl-Schwarzschild-Str. 1, D-85741, Garching, Germany}



\begin{abstract}
The cores of pulsars are expected to become superconducting soon after birth. The transition to type-II superconductivity is associated with the bunching of magnetic field lines into discrete superconducting flux tubes which possess enormous tension. The coupling of the crust to the flux tubes implies the existence of huge tangential magnetic fields at the crust-core interface. We show that the transition to superconductivity triggers a highly non-linear response in the Hall drift of the crustal magnetic field, an effect which was neglected in previous numerical modelling. We argue that at the time of the phase transition giant Hall waves are launched from the crust-core interface toward the surface.  Our models show that if the crust contains a multipolar magnetic field $\sim 10^{13}$~G, the amplitude of the Hall waves is $\sim 10^{15}$~G. The elastic deformation of the lattice is included in our models, which allows us to track the time-dependent shear stresses everywhere in the crust. The simulations indicate that the Hall waves may be strong enough to break the crust, and could cause star quakes which trigger rotation glitches and changes in the radio pulse profile. The Hall waves also couple to slow magnetospheric changes which cause anomalous braking indices. The emission of the giant Hall waves from the crust-core interface facilitates fast flux expulsion from the superconducting core, provided that the flux tubes in the core are themselves sufficiently mobile. For all of the flux tube mobility prescriptions implemented in this work, the core approaches the Meissner state with $B=0$ at late times.
\end{abstract}


\section{Introduction}
Theorists have been interested in the internal magnetic field evolution of pulsars since the early work of \cite{pacini_rotating_1968,Ginzburg_1969,baym_superfluidity_1969}. The problem is important because the magnetic field directly couples the neutron star interior to the magnetosphere. Therefore, signatures of the internal evolution may be directly observed in the pulsar emission and spin-down. 

Early numerical modelling of magnetic field evolution in neutron star interiors neglected the core entirely, and considered the evolution of crust-confined magnetic fields due to Hall drift and Ohmic diffusion \citep{hollerbach_hall_2004,pons_magneto--thermal_2009,vigano_new_2012,gourgouliatos_hall_2014,marchant_stability_2014}. Other studies attempted to model the magnetic field evolution in the core, however the problem is complicated by the effects of proton superconductivity \citep{ruderman_neutron_1998, ruderman_biography_2004, elfritz_simulated_2016, bransgrove_magnetic_2018}. 

Nuclear superfluidity in neutron star interiors was predicted by \cite{migdal_superfluidity_1959}, eight years before the discovery of the first pulsar. It is now generally accepted that below a critical temperature $T_c \sim 10^{9}-10^{10}$~K, conditions in the outer core are suitable for the Cooper pairing of protons and the formation of a type-II superconductor with critical field $H_\text{c1}\sim 10^{15}$~G \citep{baym_superfluidity_1969}. An important feature of type-II superconductors is the quantization of the magnetic field into discrete superconducting flux tubes, each of which possess a quantum of flux $\phi_0 = hc/(2e)$. The bunching of magnetic field lines into discrete flux tubes dramatically modifies the macroscopically averaged Maxwell stress. In particular, the anisotropic components of the Maxwell tensor are larger than $B^2/(4\pi)$ by a factor $H_\text{c1}/B \sim 10^3$ for a typical radio pulsar \citep{jones_alignment_1975,easson_magnetohydrodynamics_1979}.

The enhanced anisotropic Maxwell stress of the superconductor plays a key role in the tension-driven motion of flux tubes, and is one of the main qualitative differences compared to the magnetohydrodynamics of normal conducting fluids \citep{easson_stress_1977, jones_neutron_1991, jones_type_2006, gusakov_2016, gusakov_2019, Rau_2020}. The works cited above feature a wide range of predictions for the mobility of the superconducting flux tubes in response to external forces. The latter could be due to the neutron superfluid vortices pushing on the flux tubes in the process of the pulsar spin-down, or could be due to the flux tubes own magnetic tension that
tends
to straighten them. In this work we do not address the uncertain microphysics of the flux tube mobility in the core. Instead we focus on the crust-core interface, and show that the continuity of the tangential magnetic stress across the interface, combined with the type-II superconductivity in the core, plays a huge role in magnetic dynamics of the crust. The continuity of tangential magnetic stress was first realized in works attempting to build global models for magnetic equilibrium \citep{akgun_2008}, but has not been applied to magnetic-field evolution inside a neutron star.

The plan of this paper is as follows. In Section~\ref{ns_model} we describe the parameters of our model neutron star. In Section~\ref{coupling} we discuss the formulation of the correct boundary condition at the crust-core interface, and the effect of enhanced flux tube stress in activating Hall drift in the neutron star crust. In Section~\ref{Numerical} we describe the equations of motion and the numerical method. In Section~\ref{models} we present numerical simulations with the correct boundary condition, and show that giant Hall waves are launched due to the coupling of the crust to the superconducting core. In Section~\ref{discussion} we discuss our results and the implications for observed pulsar phenomenology. 

\section{Neutron Star Model}
\label{ns_model}
We obtain the density profile $\rho(r)$ for our model neutron star by integrating the equations of general relativistic hydrostatic equilibrium with the SLy equation of state \citep{douchin_unified_2001}, with a central density of $\rho_c = 10^{15}$~g~cm$^{-3}$. This gives a neutron star of radius $r_* = 11.66$~km and mass $M_* = 1.4$~M$_\odot$. The crust-core boundary is located at $r_c = 10.8$~km and density $\rho_c = 1.27\times 10^{14}$~g~cm$^{-3}$.

The electrical conductivity of the crust is very important for the magnetic field evolution of the neutron star. Unfortunately this quantity is strongly temperature-dependent, which means that the detailed model of the magnetic field evolution should include the thermal evolution of the crust. Such models of the magneto-thermal evolution were constructed e.g by \cite{vigano_unifying_2013}; they require special-purpose codes. Here we take a different approach and assume a constant internal temperature of $T=10^8$~K, appropriate for a neutron star in the first $\sim 100$~kyr of its life (the timescale for the development of giant Hall waves). At this temperature, the density of the solid-liquid phase transition at the top of the crust occurs at $\rho_\text{crys} = 10^{8}$~g~cm$^{-3}$ and $r_\text{crys} = 11.66$~km, our chosen surface cut-off. The assumption of constant temperature should be lifted in future investigations, in order to obtain more reliable conclusions.

We set the composition of the deep crust and the shallow crust according to the models of \cite{potekhin_analytical_2013} and \cite{Pearson_2011} respectively. The electron density profile $n_e (r)$ varies from $n_e\approx 2.5\times 10^{36}$~cm$^{-3}$ at the crust-core interface, to $n_e\approx 3\times 10^{31}$~cm$^{-3}$ at the surface. We set the electrical conductivity profile $\sigma(r)$ according to electron-phonon scattering \citep{baiko_thermal_1996}. We further simplify the problem by assuming isotropic conductivity, neglecting the influence of the local magnetic field (the actual conductivity tensor becomes anisotropic in the presence of strong $B$). At the temperature $T\approx 10^8$~K the conductivity varies from $\sigma\approx 10^{25}$~s$^{-1}$ at the crust-core interface, to $\sigma\approx 7\times 10^{19}$~s$^{-1}$ at the surface. The presence of lattice impurities would reduce the electrical conductivity \citep{flowers_evolution_1977, jones_first-principles_2001}. Therefore, phonon scattering provides an upper bound on $\sigma$.  

The material strength of the crust is determined by its elastic shear modulus profile $\mu(r)$. It is controlled by the Coulomb energy of the crystal lattice $\mu\approx 0.12 n_i (Ze)^2/a$, where $a\sim n_i^{1/3}$ is the mean distance between ions (nuclei) of density $n_i$ \citep{ruderman_crystallization_1968, strohmayer_shear_1991}. In the absence of impurities, the lattice fails mechanically when it is stressed beyond its critical strain $\epsilon_\text{crit}\sim 0.1$ \citep{horowitz_breaking_2009}, or even greater if the deep crust contains a layer of nuclear pasta \citep{caplan_2018}. Yielding may occur at a smaller critical strain, if the crust is polycrystalline \citep{kobyakov_elastic_2015, Baiko_2018}, if the braking stress is built up on long timescale or if a significant concentration of impurities is present in the lattice.

\section{Activation of Hall Drift by Superconducting Phase Transition}
\label{coupling}

\begin{figure}[t!]
\centering
\includegraphics[width=0.47\textwidth]{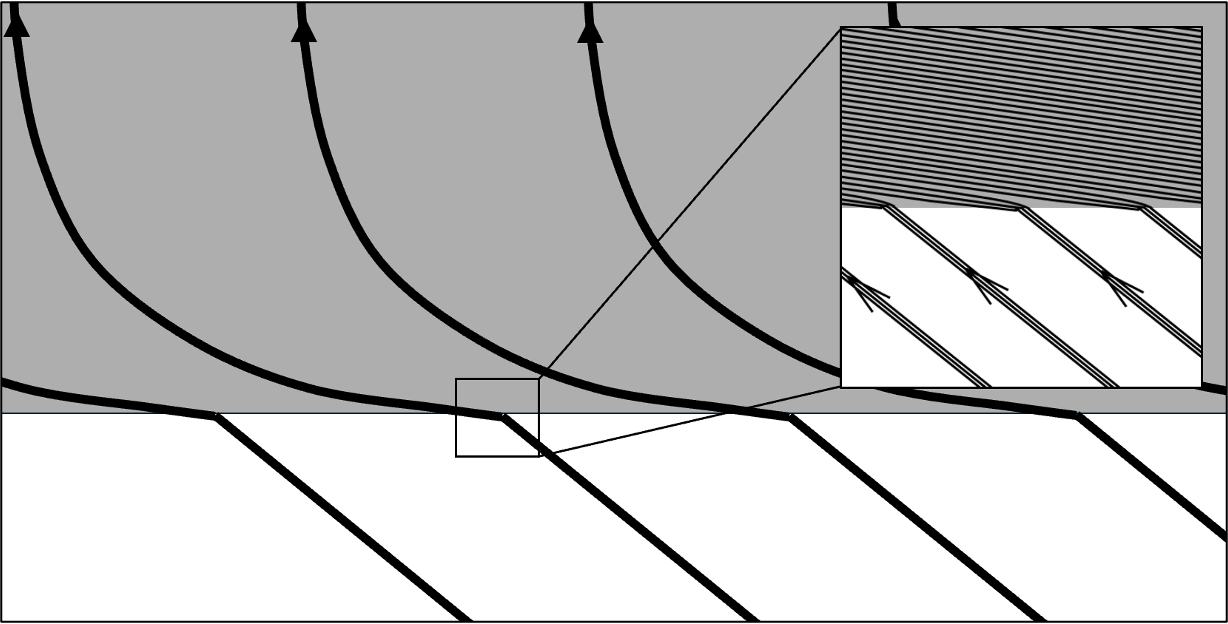} 
    \caption{Structure of magnetic field at the crust-core interface. The grey region shows the deep crust, and the white region indicates the type-II superconducting core. Thick black curves show the macroscopic average magnetic field $\vec{B}$. The zoom-in show the microscopic magnetic field (thin black curves):  individual superconducting flux tubes merge to form a ``flux sheet" at the crust-core interface. }
    \label{0}
\end{figure}
\subsection{Continuity of Stress at the Crust-Core Interface}
Consider a plane-parallel crust with the vertical coordinate $0<z<h$, where $z=0$ and $z=h$ correspond to the bottom and the top of the crust, respectively. Consider a magnetic field $\vec{B}(z)$ inside the crust that joins to an inclined array of flux tubes below the crust. Choose the coordinates so that the flux tubes just below the crust are located in the $x-z$ plane, with 
 \begin{equation}
    \vec{B}=B\left[
    - \hat{x}\,\cos \bar{\theta} + \hat{z}\,\sin \bar{\theta} 
    \right. ].
 \end{equation}
Here $\bar{\theta}$ is the angle between the flux tubes and the crust-core interface, and $\vec{B}$ represents the macroscopically-averaged magnetic field below the interface. 
 
The macroscopically-averaged tangential Maxwell stress on the core side of the interface is given by
 \begin{equation}
     T_{xz}(0^-)=\frac{H_x B_z}{4\pi } = - \frac{H B}{4\pi }
     \,\sin \bar{\theta} \cos \bar{\theta},
     \label{maxwell}
\end{equation}
where $H\sim 10^{15}\hbox{G}$ is the critical magnetic field inside a flux tube \citep{easson_stress_1977}. Discontinuity of the Maxwell stress would imply a singular current sheet at $z=0^+$. This would imply an infinite Ohmic dissipation power in the resistive medium, which is unphysical. Therefore, the tangential components of the Maxwell stress $T_{xz}$, $T_{yz}$ must be continuous across the interface. The same conclusion was reached by \cite{akgun_2008}, and later implemented in the equilibrium models of \cite{lander_2012, Henriksson_2013, lander_2013, lander_2014}. Matching Eq.~\ref{maxwell} to $B_x B_z/(4\pi)$ on the crust side of the interface, and noting that $B_z(0^+)= B_z(0^-)$ by conservation of vertical flux, we obtain 
 \begin{equation}
     B_x(0^+) = -H \,\cos \bar{\theta} = H_x(0^{-})
 \end{equation}
 Given that $H/B\gtrsim 10^3$ for typical pulsars, this is a remarkable enhancement compared to the $B_x(0^+) = B_x(0^-)$ matching condition which is satisfied before the phase transition, and was incorrectly used in the models of 
 \cite{bransgrove_magnetic_2018}
and \cite{elfritz_simulated_2016}.

 One of the dramatic consequences of this boundary condition is that it enables the field lines to slide much more easily along the crust-core interface, thus rapidly pumping the horizontal flux into the crust and at the same time enabling the vertical flux to evacuate the core. This effect is best demonstrated in a 1-dimensional model presented in the next two subsections.
 
\subsection{1D Steady State Magnetic Field Evolution: Crust}
Let us assume again the plane-parallel geometry. As an extra constraint, assume also that while the magnetic field has all three components, they depend only on the $z$-coordinate. The vertical field $B_z$ is constant, but the horizontal field changes with time. It is convenient to define a complex field 
 \begin{equation}
     \tilde{B}(z,t)=B_x+i B_y.
 \end{equation}
 It's evolution is given by the following equation \citep{cumming_magnetic_2004}
 \begin{equation}
     {\partial \tilde{B}\over \partial t}={\partial\over \partial z} \left[(\eta-i B_z D){\partial \tilde{B}\over \partial z}\right],
     \label{motion}
\end{equation}
where $\eta$ is the magnetic diffusivity and $D=c/(4\pi n_e e)$, where $n_e$ is the number density of conducting electrons and $e$ is the absolute value of the electron charge. The evolution equation here neglects the advection term due to 
time-dependent
deformation of the crystal lattice by time-dependent magnetic stress. This term, introduced and shown to be important by \cite{cumming_magnetic_2004}, is included in our multidimensional numerical models of the later section.
 
Let us now look for a stationary solution, by setting $\partial {\tilde B}/\partial t=0$, and by imposing the 
customary
boundary condition $\tilde{B}(h)=0$ 
(it
insures that there is no horizontal magnetic stress at the top of the crust). This gives 
 \begin{equation}
    \begin{split}
     {\partial \tilde{B}\over \partial z}=-\tilde{B}(0_+)[\eta(z)- & i B_z D(z) ]^{-1} \\ & \times \left[\int_0^h {dz_1\over \eta(z_1)-iB_z D(z_1)}\right]^{-1}.
     \end{split}
 \end{equation}
Using 
$\tilde{B}(0^+)=-H \cos \bar{\theta}$,
we obtain the following stationary field in the crust
 \begin{equation}
    \begin{split}
     \tilde{B}_{\rm st}(z)=-H 
     \,\cos \bar{\theta}
     & \left[\int_z^h {dz_2\over \eta(z_2)-iB_z D(z_2)}\right] \\ & \times \left[\int_0^h {dz_1\over \eta(z_1)-iB_z D(z_1)}\right]^{-1}
    \end{split}
 \end{equation}
This steady-state solution features horizontal currents which 
advect
magnetic field lines in the crust. 
The horizontal advection of  field in the crust
has to be matched by the motion of flux tubes in the core; this is crucial for the evolution of the core magnetic field,
as
discussed in the next subsection.

\subsection{1D Steady-State Magnetic Field Evolution: Core}
The motion of the core flux tubes can be obtained by matching the horizontal electric field across the crust-core interface. It is convenient to introduce $\tilde{E}(z)=E_x+i E_y$ in the crust. It is related to the magnetic field through the following equation:
\begin{equation}
     \tilde{E}=\left(B_z D+i\eta\right){\partial \tilde{B}\over \partial z}.
\end{equation}
The first term represents the Hall term, and the second term represents the Ohmic resistivity. 

The electric field in the core (treated as an ideal conductor with negligible Hall drifts) is given by
\begin{equation}
\label{eq:Ec}
    \vec{E} =-\frac{\vec{v}\times \vec{B}}{c},
\end{equation}
where $\vec{v}$ is the velocity of the flux tubes. We remind the reader that below the crust-core interface $\vec{E}$ and $\vec{B}$ represent the macroscopically averaged fields. Defining $\tilde{v}=v_x+iv_y$, 
we can write Equation~(\ref{eq:Ec}) as
\begin{equation}
    \tilde{E} = i\tilde{v}B_z .
\end{equation}

Matching the horizontal electric field across the crust-core interface, we get
\begin{equation}
    \tilde{v}(t)=(\eta/B_z-i D)~{\partial \tilde{B}(t,z=0^+)\over \partial z}.
\end{equation}
In the steady state, the velocity of the flux tubes is 
\begin{equation}
    \tilde{v}={H\over B}
    \,\cot \bar{\theta}
    \left[\int_0^h {dz_1\over \eta(z_1)-iB_z D(z_1)}\right]^{-1},
    \label{v}
\end{equation}
which is faster than the velocity in \cite{bransgrove_magnetic_2018} (Eq.~72 therein) by a factor of order $H/B \gtrsim 10^3$. 

As seen in the full time-dependent model described below, the horizontal displacement of the flux tubes in the core develops on a timescale far shorter than the resistive timescale in the crust above the interface where the magnetic field remains frozen. This leads to horizontal shearing of the magnetic field lines in the lower crust. The boundary condition $B_x(0^+) = H_x(0^-)$ speeds up the pumping of horizontal magnetic flux into the lower crust by the factor of $\sim H/B$ compared to previous models that used $B_x(0^+) = B_x(0^-)$. This dramatically boosts the nonlinear Hall evolution and elastic deformation of the crust.

\section{Description of the Full Model}  
In this section we describe the equations of motion which are solved in our axisymmetric numerical simulations. Sections~\ref{crust} and \ref{elasticity} describe the magnetic field evolution in the crust, and Section~\ref{Core} describes our prescriptions for magnetic field evolution in the core.

\label{Numerical}
\subsection{Crust}
\label{crust}
This section follows exactly the treatment of the field evolution in the crust as described in \cite{bransgrove_magnetic_2018}. In the neutron star crust the positively charged ions (nuclei) are locked into a crystal lattice, and the magnetic field is advected by degenerate electrons. The magnetic field evolves according to the equation 
\begin{equation}
    \frac{\partial\vec{B}}{\partial t} = \nabla \times \left( \vec{v}_e \times \vec{B} \right) - \nabla\times \left(\eta \nabla \times \vec{B} \right)
        \label{hall_induction}
\end{equation}
where $\vec{B}$ is the magnetic field, $\vec{v}_e$ the velocity of the electron fluid, $\eta=c^2/(4\pi\sigma)$ the ohmic diffusivity, with $\sigma$ the electrical conductivity \citep{jones_neutron_1987,goldreich_magnetic_1992}.

The velocity of electron fluid may be written as 
\begin{equation}
    \vec{v}_e = \vec{v}_\text{hall} + \vec{v}_i = -\frac{c}{4\pi n_e e}\nabla\times\vec{B} + \vec{v}_i
        \label{velocity}
\end{equation}
where $n_e$ is the electron density, and $e$ the (positive) electron charge. The Hall drift velocity $\vec{v}_\text{hall} = \vec{v}_e - \vec{v}_i = -\vec{j}/(n_e e)$ is determined by Ampere's law $\vec{j}=(c/4\pi)\nabla\times \vec{B}$. The velocity of the ion lattice $\vec{v}_i$ is due to the deformation of the solid crust in response to the applied $\vec{j}\times\vec{B}/c$ force.

 For small strains ($\epsilon\ll 1$), the Lagrangian displacement of the crustal lattice $\boldsymbol{\xi}(\boldsymbol{r})$ is described by linear elasticity \citep{landau_theory_1970}. We assume that strong radial stratification allows only horizontal deformations ($\xi_r = 0$) and the deformations are incompressible ($\nabla\cdot \vec{\xi}=0$), since the  hydrostatic pressure far exceeds the Coulomb energy density of the lattice.\footnote{In axisymmetry, this implies $\vec{\xi}=\xi_\phi(r,\theta)\vec{\hat{\phi}}$.} 
Since the forcing of the crust occurs on Hall drift timescales (much slower than the elastic wave crossing time), we may neglect the inertia of the crust, and assume that the system evolves through a sequence of magneto-elastic equilibria. The equilibrium is defined by
\begin{equation}
    0 = (\nabla\mu \cdot \nabla)\vec{\xi} - (\vec{\xi}\cdot\nabla)\nabla\mu + \mu\nabla^2\vec{\xi} + \frac{\vec{j}\times\vec{B}}{c},
        \label{force_balance}
\end{equation}
where $\mu$ is the shear modulus of the crust, and the form of the elastic restoring force follows from Hooke's law \citep{landau_theory_1970,bransgrove_magnetic_2018}. Instead of dealing with the above constraint equation, we use the relaxation method of \cite{bransgrove_magnetic_2018} to find $\vec{v}_i = \dot{\vec{\xi}}$, and enforce Eq.~\ref{force_balance} dynamically. 

\subsection{The Importance of Including the Elastic Deformation of the Crust}
\label{elasticity}
One of the key features of our model is the self-consistent elastic deformation of the crust. It is essential for this problem because the feedback of the elastic deformation on the magnetic field modifies the amplitude and propagation of Hall waves \citep{cumming_magnetic_2004}. It also allows us to track the time-dependent elastic strains everywhere in the crust, 
diagnose the location of likely crustal failures, and the possibility of star quakes.

Many previous studies neglected the magneto-elastic equilibrium (Eq.~\ref{force_balance}) and incorrectly assumed stress balance \citep{perna_unified_2011, lander_2019, dehman_2020, gourgouliatos_2021}. That is, they calculated elastic stresses by equating them to local Maxwell stresses. We emphasize that this is incorrect: In magneto-elastic equilibrium the divergence of stress (i.e. the force) is balanced, not the stress itself. Due to differing spatial gradients of the magnetic field and the elastic deformation, we find that the elastic stress is often smaller than the local Maxwell stress by several orders of magnitude for configurations which satisfy magneto-elastic equilibrium. Therefore, studies which assume stress balance likely overestimate the rate of crustal failures, and their conclusions should be taken with extreme caution. 

\subsection{Core}
\label{Core}
Neutron star cores contain degenerate nuclear matter which consists of neutrons, protons and electrons. Protons are expected to form Cooper pairs through the attractive short range component of the strong nuclear force. The properties of the superconductor are determined by its penetration depth $\lambda_p = (m_p c^2 / 4\pi n_p e^2)^{1/2}$, and coherence length $\xi_p\approx (2/\pi k_F)(\epsilon_p / \Delta_p)$ where $k_F$ is the proton Fermi wavevector, $\epsilon_p$ the proton Fermi energy, and $\Delta_p$ the superconducting energy gap \citep{baym_superfluidity_1969}. Typical densities in the outer core imply $\lambda_p/\xi_p \gg 1 $, so the superconductor is of type-II. Thus, at the time of the phase transition the magnetic field is quantized into an array of discrete superconducting flux tubes which each contain the quantum of flux $\phi_0 = hc/(2e)$ within a characteristic radius $\lambda_p$. The local magnetic field inside a flux tube is $\vec{H} \approx H_\text{c1} \vec{B}/|\vec{B}|$, with $H_\text{c1} \approx \phi_0 /(\pi \lambda_p^2)\sim 10^{15}$~G, while the field is exponentially suppressed in the volume between flux tubes \citep{easson_stress_1977}. In the neutron star core, we use $\vec{B}$ to represent the magnetic field which is spatially averaged over many flux tubes such that $|\vec{B}| = n_\Phi \phi_0$, where $n_\Phi$ is the local area density of flux tubes.

Generally, motion of the flux tube array with velocity $\vec{v}_L$ induces a spatially averaged electric field $\vec{E}=-\vec{v}_L\times \vec{B} /c$. According to Faraday's law, the magnetic field $\vec{B}$ then evolves according to the induction equation
\begin{equation}
    \frac{\partial\vec{B}}{\partial t} = \nabla \times \left( \vec{v}_L \times \vec{B} \right). 
    \label{induction}
\end{equation}
The flux tube velocity may be written as $\vec{v}_L = \vec{v}_d + \vec{v}$, where $\vec{v}_d$ is the dissipative drift of flux tubes relative to the background fluid, and $\vec{v}$ is the velocity of the background fluid due to bulk hydromagnetic flows. The flux tube drift is slow compared to the Alfv\'en crossing time of the core. Therefore, hydromagnetic flows (described by velocity $\vec{v}$) ensure that the core evolves through a sequence of ideal MHD equilibria \citep{bransgrove_magnetic_2018}, which are described below. While the equilibria are straightforward to compute, the dissipative drift that takes place on top of the equilibrium configurations depends on poorly understood microphysics of the superconducting flux tubes. To address this uncertainty, we implement a range of physically motivated prescriptions for the drift velocity $\vec{v}_d$ and explore the implications of each of them. Our experiments allow us to explore the potential impact of the phase transition on the evolution of the magnetic field in the crust (Sections~\ref{Jones} \& \ref{Vortex}). 

\subsubsection{Hydromagnetic Equilibrium}
\label{MHD}
For simplicity we assume that statification of the core suppresses radial displacements ($v_r = 0$), and the allowed horizontal motions are incompressible ($\nabla \cdot \vec{v} = 0 $). In axisymmetry, it then follows that the only allowed fluid motions are in the $\phi$ direction ($v_\theta = 0$). Therefore, hydromagnetic equilibrium requires the azimuthal force
\begin{equation}
    f_\phi = \frac{1}{4\pi}[(\vec{B}\cdot \nabla )\vec{H}]_\phi = 0,
    \label{f_phi}
\end{equation}
which is a constraint of the evolution. Rather than incorporating the above constraint equation into the code directly, we prefer to solve a relaxation problem and enforce Eq.~\ref{f_phi} dynamically \citep{yang_force-free_1986, bransgrove_magnetic_2018}.

\begin{table*}
\centering
\caption{Sample models. All models have a magnetic dipole moment with strength at the pole $B_d = 3\times 10^{12}$~G, except model D2 which has $B_d = 1.5\times 10^{13}$~G. The second column shows the assumed timescale of flux tube relaxation $\tau$ (Eq.~\ref{vd}), and the fourth column indicates the vortex-flux tube interactions: no pinning (Eq.~\ref{vd}), or strong pinning (Eq.~\ref{vf}). The fifth column indicates the initial RMS magnetic field strength $\bar{B} = \sqrt{\langle B^2\rangle }$, where the spatial average $\langle ...\rangle $ is taken over the crust volume. }
\begin{tabular*}{\textwidth}{c @{\extracolsep{\fill}} ccccccc}
\hline
Model & Initial Condition & $\tau$           & vortex-flux tube interactions & $ \bar{B} $  \\
\hline
D1   & Dipole             &    0~kyr          &  no pinning              & $2 \times 10^{12}$~G  \\
D2   & Dipole             &    0~kyr          &  no pinning              & $1 \times 10^{13}$~G  \\
M1   & Dipole + Multipole &    500~kyr        &  no pinning              & $1 \times 10^{13}$~G  \\
M2   & Dipole + Multipole &    500~kyr        & strong pinning           & $1 \times 10^{13}$~G \\
M3   & Dipole + Multipole &    500~kyr        & strong pinning           & $3 \times 10^{13}$~G \\
\hline
\end{tabular*}
\label{Sample_models}
\end{table*}

\subsubsection{Flux Tube Drift (No pinning)}
\label{Jones}

Here we describe a model for the dissipative drift of flux tubes while neglecting their pinning with magnetized neutron vortices. It is appropriate if the vortex-flux tube pinning energy is small, or if vortex-flux tube reconnection occurs very easily. The motion of flux tubes has been a subject of considerable theoretical investigation and debate. In this work, we use a simple toy model in order to study the effects of crust-core coupling. Similar to \cite{jones_type_2006} and \cite{bransgrove_magnetic_2018}, we assume that flux tubes drift due to their own self-tension force with velocity
\begin{equation}
    \vec{v}_d = \frac{\alpha}{4\pi}(\vec{B}\cdot \nabla )\vec{H}.
    \label{vd}
\end{equation}
For simplicity we neglect the density dependence of $\vec{H}$. We parameterize the drag coefficient as $\alpha\equiv 4\pi R_* / (BH\tau)$, where $\tau$ is the characteristic time for a flux tube with radius of curvature $R_*$ to dissipatively straighten. \cite{jones_type_2006} calculates $\alpha$ due to electron scattering off quasi-particles localized in the flux tube cores - this corresponds to $\tau \sim 500$~kyr. Other  studies have found far longer timescales due to electron scattering off magnetic flux localized in the tubes \citep{gusakov_2019, gusakov_2020}. In this work we focus on drag coefficients corresponding to $\tau \leq  500$~kyr, so that the flux tubes are mobile on the evolutionary timescale of the crust. We discuss the effect of $\tau\gg 500$~kyr in Section~\ref{models}.

\subsubsection{Flux Tube Drift (Strong pinning)}
\label{Vortex}

In this section we describe a simple toy model for the flux tube drift which includes the effects of vortex-flux tube interactions. The neutron vortices are magnetized by the proton entrainment effect. Therefore, the flux inside the vortices could make them into pinning sites for the flux tubes \citep{srinivasan_novel_1990, ruderman_neutron_1998, jones_type_2006}, and could result in advection of magnetic flux by vortex motion. As the neutron star spins down due to external torques, the superfluid neutron vortices move outward with velocity 
\begin{equation}
    \vec{v}_\perp = -\frac{r_\perp \dot{\Omega}_n}{2\Omega_n}\vec{\hat{r}}_\perp,
    \label{v_perp}
\end{equation}
where $\Omega_n$ is the angular frequency of the neutron superfluid, and $r_\perp$ is the cylindrical radius.
The flux tubes are also free to slide dissipatively along the neutron vortices which have local orientation $\vec{\hat{e}}_n$.  However, if the force of the flux tube array on the neutron vortices becomes too great, it is energetically favorable for the flux tubes to cut through the vortices, and reconnect \citep{ruderman_neutron_1998,jones_type_2006,gugercinoglu_microscopic_2016}. We capture the basic dynamics of this scenario in a 
toy model for the flux tube velocity:
\begin{equation}
    \vec{v}_d = 
    \begin{cases}
      \vec{v}_\perp + \left[\frac{\alpha}{4\pi}(\vec{B}\cdot\nabla)\vec{H}\cdot\vec{\hat{e}}_n\right]\vec{\hat{e}}_n & \text{(advection)}\\
      \vec{v}_\text{max} + \frac{\alpha}{4\pi}(\vec{B}\cdot\nabla)\vec{H} & \text{(cut-through),}
    \end{cases}
    \label{vf}
\end{equation}
where $\vec{v}_\text{max}$
is the maximum velocity of a flux tube being advected by neutron vortices before reconnection occurs \citep{ruderman_neutron_1998, jones_type_2006, gugercinoglu_microscopic_2016, bransgrove_magnetic_2018}. We assume that the neutron vortices are perfectly rigid, and aligned with the polar axis ($\vec{\hat{e}}_n = \vec{\hat{z}}$). Further details of the model are discussed in \cite{bransgrove_magnetic_2018}. Numerical simulations of quantum fluids suggest that vortex-flux tube reconnection occurs easily if the arrays are tangled on microscopic scales 
\citep{drummond_2017,drummond_2018,thong_2023}. Note that in the limit of a negligible threshold for vortex-flux tube reconnection $|\vec{v}_\text{max}|\longrightarrow 0$, and the flux tube drift prescription of Eq.~\ref{vf} reduces to that of Eq.~\ref{vd}.

\section{Sample Models}
\label{models}

\begin{figure*}[t!]
\centering
\includegraphics[width=1.05\textwidth]{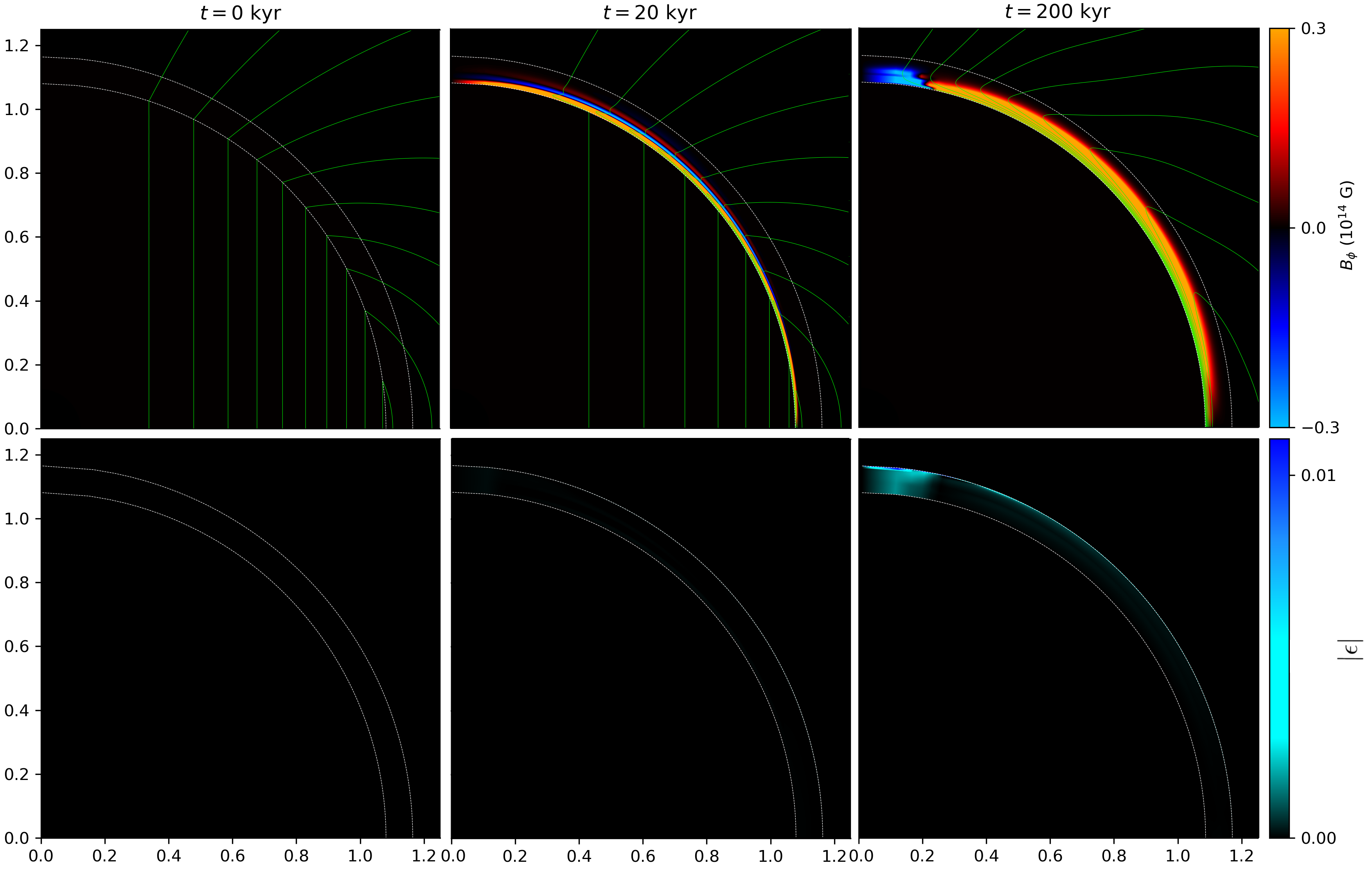} 
\caption{Magnetic field evolution of Model D2. Top row: Snapshots of the magnetic field configuration at $t=0$~kyr, $t=20$~kyr, and $t=200$~kyr. Green curves show poloidal magnetic field lines, and color shows $B_\phi$ in units of $10^{14}$~G. Bottom row: Snapshots of the von Mises strain $|\epsilon | = \sqrt{\frac{1}{2}\epsilon_{ij}\epsilon_{ij}}$ at the same times. The crust-core interface and the surface are indicated by the inner and outer dashed white curves, respectively. The axis show distance in units of $10^6$~cm. An animated version of this figure, showing the magnetic field evolution from $t=0$~kyr to $200$~kyr, is available in the online article.}
\label{1}
\end{figure*}

\begin{figure*}[t!]
\centering
\includegraphics[width=1.05\textwidth]{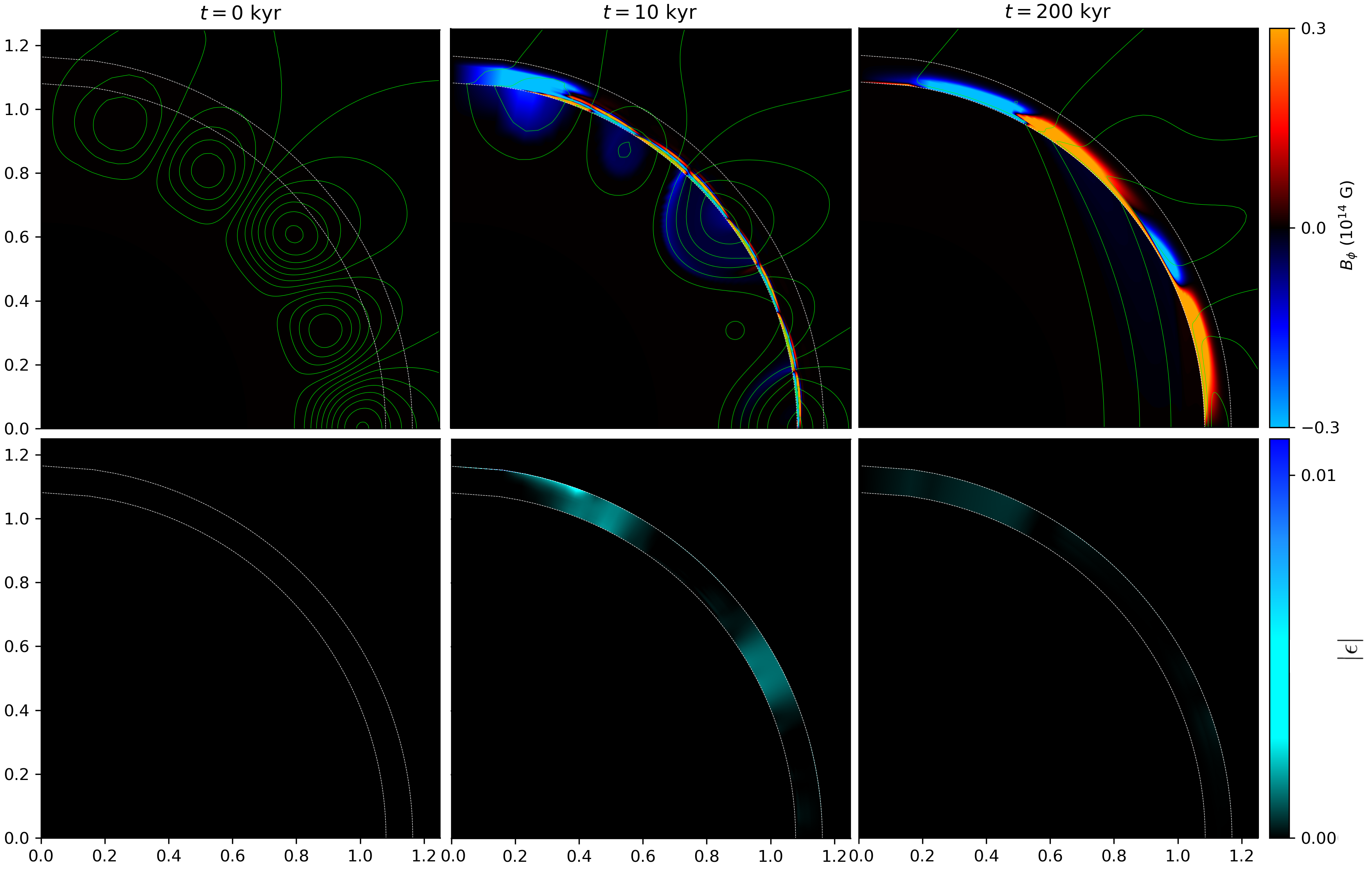} 
    \caption{Same as Fig.~\ref{1}, but for Model M1. In the core, color shows $H_\phi$ in units of $10^{14}$~G. An animated version of this figure, showing the magnetic field evolution from $t=0$~kyr to $200$~kyr, is available in the online article.}
    \label{2}
\end{figure*}

\begin{figure*}[t!]
\centering
\includegraphics[width=1.05\textwidth]{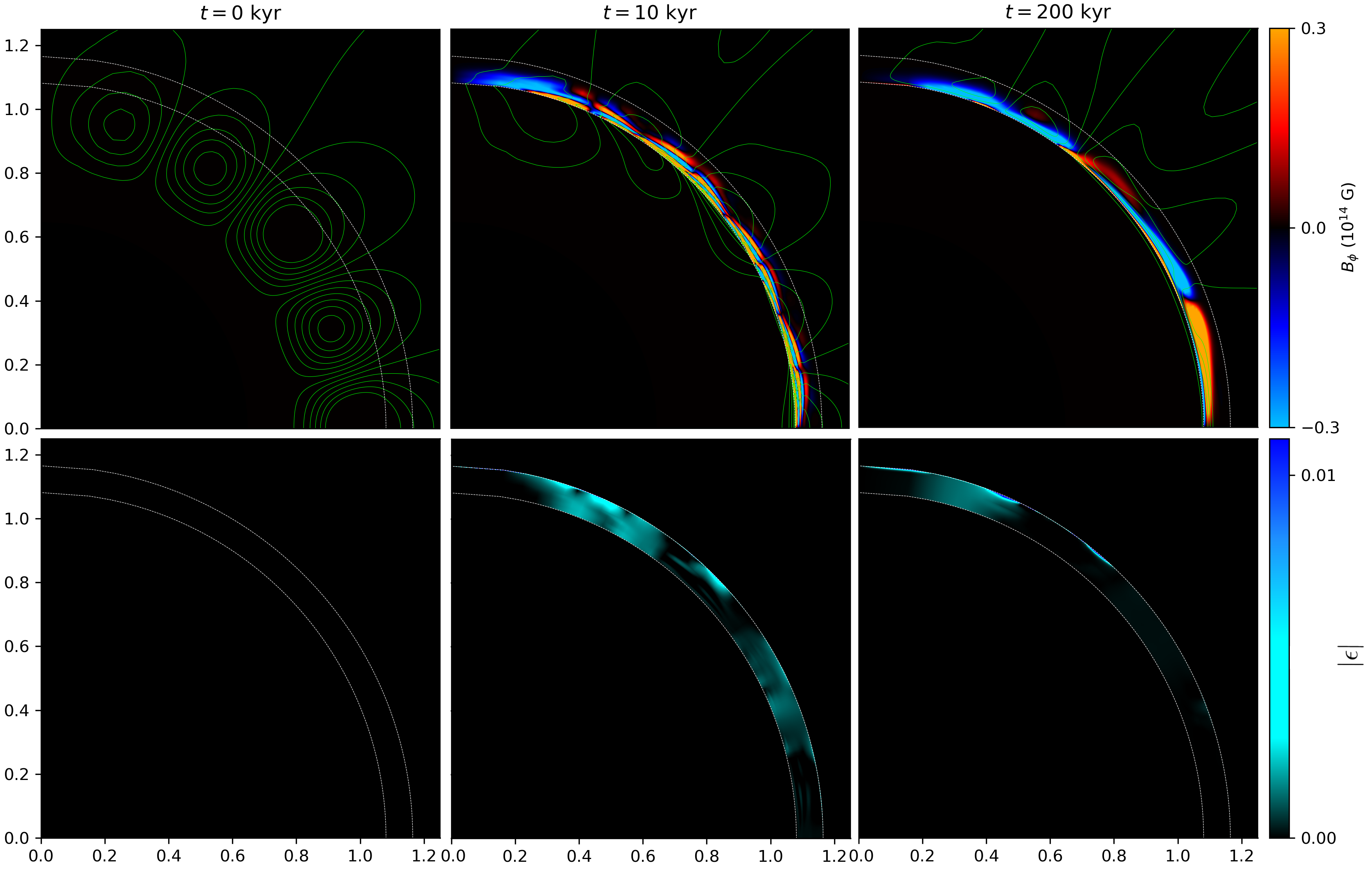} 
    \caption{Same as Fig.~\ref{1}, but for Model M3. An animated version of this figure, showing the magnetic field evolution from $t=0$~kyr to $200$~kyr, is available in the online article.}
    \label{3}
\end{figure*}

We present five sample models, which cover a range of initial magnetic fields and different models of superconducting flux tube drift in the core. The parameters of the models are summarised in Table~\ref{Sample_models}. The dynamics of the magnetic field is fascinating, and we provide movies that illustrate the field evolution corresponding to Figures~\ref{1}, \ref{2}, and \ref{3}. In addition, we focus on the following quantities: (1) The von Mises strain of the solid crust $|\epsilon|= \sqrt{\frac{1}{2}\epsilon_{ij}\epsilon_{ij}}$ and its maximum value at a given time $|\epsilon |_\text{max}$. The latter diagnoses the possibility of star quakes, and is plotted as a function of time in Fig.~\ref{n}, (2) The braking index $n=\nu\ddot{\nu}/\dot{\nu}^2$, where $\nu$, $\dot{\nu}$, and $\ddot{\nu}$ are the pulsar rotation frequency and it's time derivatives. The braking index is commonly measured by pulsar observers, and is plotted in Fig.~\ref{n}.

We solve the equations of motion described in Section~\ref{Numerical} on a spherical ($r,\theta$) grid which is uniformly spaced in $r$ and $\cos\theta$. For Models D1 and D2 the domain covers the crust $r \in [10.8, 11.66]$~km and $\theta \in [0,\pi]$ with $N_r \times N_\theta = 100 \times 400$ grid cells. For all other models the domain covers the outer core and crust $r \in [6.5, 11.66]$~km and $\theta \in [0,\pi]$ with $N_r \times N_\theta = 600 \times 200$ grid cells. We use 4th order Kreiss-Oliger dissipation to suppress high frequency numerical noise which develops at current sheets due to our explicit finite difference scheme \citep{KO_1973}. Further details of the numerical method are described in \cite{bransgrove_magnetic_2018}. 

Ohmic decay of electric currents in the closed magnetosphere around the neutron star is fast compared to the internal magnetic evolution studied in this paper. Therefore, outside the neutron star surface, $r>R_\star$, we assume a vacuum magnetosphere so that the magnetic field at $R_\star$ matches onto a potential field ($\nabla\times\vec{B}=0$) at each time step \citep{bransgrove_magnetic_2018}. We assume that the phase transition happens at $t=0$, so that the entire core is superconducting throughout the simulation. Therefore, at each time step we enforce the continuity of stress at the crust-core interface, as discussed in Section~\ref{coupling}, $B_r(r_c^+) = B_r(r_c^-)$, $B_\theta(r_c^+) = H_\theta(r_c^-)$, and $B_\phi(r_c^+) = H_\phi(r_c^-)$, where $r_c$ is the radius of the crust-core interface. All of the initial magnetic fields satisfy $\nabla\times\vec{B}=0$ in the crust, so that the resulting evolution is entirely due to the crust-core coupling. We choose initial magnetic fields with $B\ll H$ similar to real pulsars, so that the magnetic stress in the core is much larger than the magnetic stress in the crust. This ensures that the Hall waves have a very large amplitude. For magnetars with $B\sim H$, the magnetic stress in the crust and the core is similar, and the effect of the phase transition is much less dramatic.

\subsection{Infinitely fast flux tube relaxation ($\tau = 0$)}
Models D1 and D2 assume the simplest possible configuration: an initially dipolar magnetic field in the crust which matches onto vertical flux tubes in the core.  We assume infinitely fast flux tube relaxation in the core, so that the flux tubes always remain in tension equilibrium (perfectly straight with $\vec{B} = B \hat{z}$), and we neglect vortex-flux tube interactions. With these simplifying assumptions, the boundary condition at the crust-core interface is $B_\theta(r_c^+) = H_\theta(r_c^-) = H_r(r_c^-)\tan\theta$, where $H_r(r_c^-) = B_r(r_c^+)H_{c1}/B$ by continuity of $B_r$. Equatorial symmetry ensures that the core magnetic field remains untwisted, and hydromagnetic equilibrium ($f_\phi = 0$) is satisfied trivially ($B_\phi = H_\phi = 0$ in the core) [\cite{bransgrove_magnetic_2018}, Section~\ref{MHD}]. By using these boundary conditions, we avoid directly solving the magnetic field evolution in the core for models D1 and D2. 

The initial magnetic field strength in model D1, $\bar{B}\sim 10^{12}$~G, implies the
Hall Reynolds number $R_\text{H}\equiv \sigma \bar{B}/(n_e e c)\sim 0.2-0.3 <1$. Since the flux tube relaxation is instant in this model, the drift velocity of magnetic field lines is limited by the rate of Ohmic diffusion near the crust-core interface. The strong curvature of poloidal field lines near the interface produces a powerful current sheet $j_\phi$. A quasi-steady drift is established with the current sheet occupying the deep part of the crust with scale-height $h\sim 3\times 10^4$~cm where $\sigma$ is large. The Ohmic drift velocity of the field lines can be estimated using  Eq.~\ref{v} with $D=0$, which corresponds to $R_{\rm H}\ll 1$.
It gives the characteristic timescale for the magnetic flux to be expelled from the core
\begin{equation}
    \begin{split}
         t_d & \sim 
         \frac{B}{H_{c1}}\frac{hl}{\eta}
         \,\tan \bar{\theta}
         \\ 
         &\sim 500~
         B_{12}
         \left( \frac{h}{3\times 10^4~\text{cm}}\right)\left(\frac{l}{\pi r_*}\right)\left( \frac{\sigma}{ 10^{25}~ \text{s}^{-1}} \right)~\text{kyr},
    \end{split}
     \label{t_d}
\end{equation}
where $l$ is the length along the crust-core interface, and we took the angle $\bar{\theta}\sim \pi /4$ and $B=10^{12}$~G. Note that the drift timescale Eq.~\ref{t_d} is at least $\sim 10^3$ times faster than the timescale of \cite{bransgrove_magnetic_2018} for the same electrical conductivity [Eq.~(73) therein]. 

The flux tubes drift radially outward until they have entirely entered the crust (this happens when the flux tube reaches the crust in the equatorial plane). After $\sim 200$~kyr most of the flux tubes have entered the crust, and the core is approaching the Meissner state with $B=0$. We emphasise that the rapid expulsion of flux from the core is due to the enormous tension of the flux tubes, which is communicated to the crustal magnetic field through the boundary conditions discussed in Section~\ref{coupling}. This effect was neglected in the models of \cite{bransgrove_magnetic_2018} and \cite{elfritz_simulated_2016}.

\begin{figure*}[t!]
\centering
\includegraphics[height=0.36\textwidth]{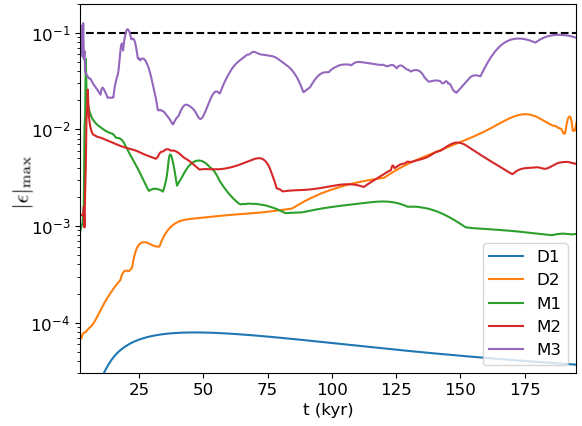} 
\includegraphics[height=0.36\textwidth]{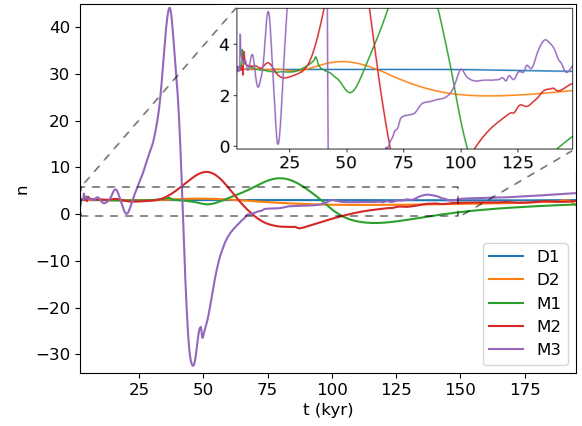} 
\caption{Left: Evolution of the maximum elastic strain in the crust $|\epsilon |_\text{max}$. The dashed black line corresponds to the breaking strain of the crust $\epsilon_\text{crit}=0.1$. Right: Evolution of the pulsar braking index $n=\nu\ddot{\nu}/\dot{\nu}^2$. Inset shows oscillations around $n=3$. }
\label{n}
\end{figure*}

Model D2 has a stronger initial magnetic field $\bar{B} = 10^{13}$~G, and the initial Reynolds number $R_\text{H}\gtrsim 1$ ensures that the Hall effect plays a significant role in the magnetic field evolution. When the simulation begins, a powerful current sheet forms at the crust-core interface, resulting in a similar radial drift of the flux tubes as Model D1. However, the larger $R_\text{H}$ results in a broad spectrum of giant Hall waves (whistler waves) being launched from the crust-core interface. The high frequency whistlers have small amplitude $B_\phi/B_p\ll 1$ ($B_p$ is the poloidal background field), and propagate along the background poloidal field \citep{goldreich_magnetic_1992}, quickly reaching the surface of the neutron star, where they are reflected back into the crust. However, the low frequency whistlers with typical wavelength $\lambda\sim 3\times 10^4$~cm are highly non-linear with amplitudes $B_\phi / B_p \sim 10^2$ ($B_\phi \sim 10^{15}$~G). The non-linear whistler waves significantly deform the background field as they spread into the outer crust where the density scale-height is smaller than their wavelength. They then behave as oscillating drift waves in $B_\phi$ [e.g. \cite{vigano_new_2012}]. The waves propagate in the $\nabla n_e \times \vec{B}_\phi \propto \hat{\vec{{\theta}}}$ direction around the circumference of the crust. The whistler waves and the drift waves couple to oscillations in the dipole magnetic field which results in large-amplitude variation in braking index $n$ of the pulsar (Fig.~\ref{n}). We note that for a fixed magnetic dipole configuration, the theory predicts $n=3$ regardless of the dipole orientation \citep{Contopoulos_1999,Gruzinov_2005,spitkovsky_time-dependent_2006}. However, braking indices $n\neq 3$ are commonly observed in middle-aged pulsars [e.g. \cite{hobbs_analysis_2010,zhang_2012,Parthasarathy_2019,Parthasarathy_2020, Lower_2021,Lower_2023}], and in this work we refer to them as anomalous. The variations plotted in Fig. 5 are in a qualitative agreement with the observational picture.

\subsection{Moderate flux tube relaxation (finite $\tau$)}
Models M1, M2, and M3 assume a dipole magnetic field with a dominant multipole component (Table~\ref{Sample_models}). This configuration is motivated by the dynamo mechanism which generates strong small-scale fields. As a result of the small-scale loops, we do not evolve the magnetic field in the entire core. Model M1 assumes no vortex-flux tube interactions, so the evolution of the core magnetic field is determined by Eq.~\ref{vd}. This corresponds to small pinning energies, or a negligible threshold for vortex-flux tube reconnection \citep{drummond_2017, drummond_2018,thong_2023}. We use the relaxation method to enforce hydromagnetic equilibrium ($f_\phi = 0$). Models M2 and M3 assume strong vortex-flux tube interactions, so the evolution of the core magnetic field is determined by Eq.~\ref{vf}. In Models M2 and M3 we assume that the vortices prevent the flux tubes from relaxing to hydromagnetic equilibrium in the azimuthal direction. Therefore, we set $v_\phi = 0$ (in general this results in $f_\phi \neq 0$). The models M1, M2, and M3 have initial magnetic fields $\bar{B}\sim 1-3\times 10^{13}$~G, so that the Hall Reynolds number $R_\text{H}\gg 1$ and the response of the crust is strongly dominated by the Hall effect.

In Model M1 the flux tubes drift due to their own self-tension force (Eq.~\ref{vd}). Flux tubes minimize their energy by increasing their radius of curvature, and shortening their length. Thus, as the flux tubes relax and straighten, poloidal field lines develop strong curvature at the crust-core interface. This produces a powerful current $j_\phi$ in the deep crust,  which shears field lines in the azimuthal direction due to the Hall effect. The strong azimuthal shear of fields in the deep crust twists flux tubes, and results in non-zero toroidal fields in the core ($H_\phi \neq 0$, Fig~\ref{2} middle panel). Currents generated by toroidal fields in the deep crust advect poloidal field lines in the $\hat{\theta}$ direction, pushing together flux tubes of opposite polarity, and forcing reconnection at the crust-core interface at $t\gtrsim 20$~kyr. After $\sim 50$~kyr, all of the flux tube loops have either entered the crust, or reconnected with neighbouring loops (Fig.~\ref{2}, last panel). The flux tubes which remain in the core extend from north to south hemisphere, and are untwisted ($H_\phi =  0$, Fig~\ref{2} last panel) due to equatorial symmetry. The flux tubes then slowly drift out of the core and enter the crust, in a manner similar to Models D1 and D2. 

In Models M2 and M3 the flux tubes are advected by superfluid neutron vortices, which move outward due to the pulsar spin-down (Eq.~\ref{vf}). Since we assume rapid rotation of the neutron star (initial period $P=10$~ms), the neutron vortices significantly re-arrange the flux tubes in the core. After $\sim 10$~kyr, most of the flux has been advected into the crust (Fig.~\ref{3}). We find that models with strong vortex-flux tube interactions usually generate stronger currents at the crust-core interface, and thus produce stronger Hall waves in the crust, since the flux is actively forced out of the core. In Models M2 and M3, the core approaches the Meissner state ($B=0$) on the spin-down timescale.

In all the models M1, M2, and M3, we observe the same qualitative evolution of the magnetic field in the crust: The drift of flux tubes in the core launches a broad spectrum of Hall waves from the crust-core interface. The high frequency waves have low amplitudes $B_\phi/B_p \ll 1$, and propagate in the linear regime along the background magnetic field. The low frequency waves are highly non-linear and significantly re-arrange the poloidal background field on longer timescales. The multipolar initial conditions result in regions of ultra-strong $B_\phi$ of opposite polarity, which drift in the opposite $\hat{\theta}$ directions and converge, forming powerful current sheets and regions of localized $\vec{j}\times\vec{B}/c$. We observe strong shearing of the crust and higher elastic strains in the vicinity of these current sheets (Fig.~\ref{2} \& \ref{3}, bottom panels). The largest strains are achieved in the upper layers of the crust. Model M3 reaches $|\epsilon |\sim 0.1$, while models M1 and M2 reach $|\epsilon |\sim \text{ few}\times 10^{-2}$ in localized regions near $\rho\sim \rho_\text{crys}$ (Fig.~\ref{n}, left panel). 

The results described in this section assume that the flux tubes are mobile on the evolutionary timescale of the crust ($\tau \leq 500$~kyr). We find that increasing $\tau\gg 500$~kyr does not significantly affect the amplitude of the initial giant Hall waves. However, the core approaches the Meissner state on a much longer timescale $\propto \tau $, and the slower pumping of horizontal magnetic flux into the crust results in less dramatic Hall evolution at late times.

\section{Discussion}
\label{discussion}

Radio pulsars show significant secular changes in their spin-down and magnetospheric emission which may be related to evolution of their magnetic fields. The physics of magnetic field evolution in neutron star crusts is robust and well studied [see \cite{Pons_2019} for a review]. However, the evolution of the core magnetic field and 
its
coupling to the crust have received considerably less attention. In this work we have highlighted an important effect which was missed in previous studies: 
the
transition to type-II proton superconductivity in the core triggers a highly non-linear response in the Hall drift of the crustal magnetic field, and may be the dominant driver of crustal magnetic field evolution during the first $\sim $~Myr of a 
pulsar's
life.

We have 
shown
that radio pulsars 
tend to develop huge
tangential magnetic fields deep in their crusts, 
orders
of magnitude stronger than the surface dipole component. One important consequence of this
effect
is the possibility of magnetically induced star quakes. 
Such
star quakes were 
usually
considered 
unlikely
in 
ordinary
pulsars because their dipole magnetic fields are 
far
weaker than those of magnetars. However, our models 
show
that if the initial crustal field $\bar{B}> 10^{13}$~G, and the flux tubes are mobile
in the core, then
the elastic strain can reach $|\epsilon |~\sim 0.1$ 
in the upper crust\footnote{We used the shear modulus $\mu$ for a monocrystal bcc lattice. If the lattice is polycrystalline, it's shear modulus may be reduced by $\sim 50\%$, increasing the effective strain we measured \citep{kobyakov_elastic_2015,Baiko_2018}.} (Fig.~\ref{n}, left).  
This strain is
sufficient to cause 
a
mechanical failure,
even if the crust has the maximum strength of an ideal crystal
\citep{horowitz_breaking_2009}.

Star quakes have long been considered as a 
possible
trigger mechanism for pulsar rotation glitches \citep{link_thermally_1996}. For example, failure of the lattice could excite seismic waves which dynamically unpin superfluid vortices and trigger a rotation glitch \citep{eichler_dynamical_2010,Bransgrove_2020}. Alternatively, if a plastic flow is initiated in the deep crust the vortices could be advected away from (or toward) the rotation axis, triggering a glitch [or anti-glitch \citep{tuo_2024}]\footnote{A plastic flow could be initiated in the deep crust if the magnetic field is stronger than we assumed in our models, or if the shear modulus is significantly reduced in the nuclear pasta.}. Unpinning vortices in one part of the crust could lead to a deflagration wave of unpinnings \citep{link_2022}. The unpinning wave could propagate inward, because the vortices in the outer crust are likely to have reversed their direction \citep{levin_2023}. Observations of the Vela pulsar during its 2016 glitch and associated radio pulse disturbance strongly suggest a star quake event \citep{palfreyman_alteration_2018, Bransgrove_2020, Gug_2020}. There are also hints of star quakes in other pulsars, with sudden or gradual changes in their radio pulse profiles, often associated with glitches \citep{welterverde_2011, Keith_2013, Palfreyman_2016, Kou_2018, Liu_2021, Liu_2022, Liu_2023, jennings_2024}.

The Hall waves excited by the phase transition also couple to slow changes of the magnetosphere.
These changes
have an observable effect on the pulsar spin-down (Fig.~\ref{n}, right),
which
is often characterized by the braking index $n = \nu\ddot{\nu}/\dot{\nu}^2$. Force-free models of the magnetosphere overwhelmingly give $n=3$, regardless of the inclination angle \citep{goldreich_pulsar_1969, Contopoulos_1999, Gruzinov_2005, spitkovsky_time-dependent_2006}. Remarkably, observed pulsars have braking indices of positive and negative sign spanning the range $-10^8 \lesssim n \lesssim 10^8$ \citep{hobbs_analysis_2010, zhang_2012}. The implication from both theory and observation is that the pulsar magnetic fields must be 
strongly evolving on a timescale much shorter than the pulsar's age.
Hall drift in the crust naturally produces anomalous braking indices $n\neq 3$ because the dipole moment becomes time-dependent -- this effect was demonstrated by \cite{pons_2012} and \cite{gourgouliatos_2015}. The braking indices measured in our simulations ($|n|\lesssim 50$, Fig.~\ref{n}) are similar to the braking indices of observed pulsars younger than $\sim 200$~kyr which span $|n|\lesssim 100$ \citep{zhang_2012, Parthasarathy_2019,Parthasarathy_2020, Lower_2021}. Our simulations produce $n$ and $\ddot{\nu}$ of positive and negative sign after the first $\sim 50-100$~kyr, similar to observed pulsars \citep{hobbs_analysis_2010}.

Finally, note that we assumed a constant temperature of the neutron star and did not model 
its
thermal evolution. Cooling increases the electrical conductivity and
the
Hall Reynolds number $R_\text{H}$, and has been shown to 
enhance
the effect of Hall waves on the spin-down 
at
late times \citep{pons_2012}. Cooling also strongly reduces 
the crystallization density
$\rho_\text{crys}\propto T^3$ and the shear modulus at the surface $\mu_\text{crys}\propto T^4$, so the outer crust breaks much more easily at late times. Also note that the elastic strain distribution could be qualitatively different if the magnetic field configuration is three-dimensional, and we leave this for future work. Future studies should also address what happens when the crust fails and starts to deform plastically.

\section*{acknowledgments}
The authors thank A. Melatos, A. Philippov, P. Rau, and A. Spitkovsky for useful discussions. 
A.B. is supported by a PCTS fellowship and a Lyman Spitzer Jr. fellowship.
A.M.B. acknowledges grant support by NSF AST 2009453, NASA 21-ATP21-0056, and Simons Foundation \#446228.
Y.L.'s work on this subject is supported Simons Collaboration on Extreme Electrodynamics of Compact Sources (SCEECS) and by Simons Investigator Grant 827103.

\bibliographystyle{apj}
\bibliography{main}

\end{document}